\documentclass[twocolumn,pra,superscriptaddress,showpacs]{revtex4-1}
\usepackage{float}
\usepackage{amssymb}
\usepackage{amsmath}
\usepackage{graphicx}
\usepackage[capitalise]{cleveref}
\usepackage{epsf}
\usepackage{tikz}
\usepackage{epsfig}
\usepackage{epstopdf}
\usepackage{color}
\usepackage{color}
\usepackage{float}
\usepackage{comment}

\def\g{{\bf{g}}}

\def\g0{{\gamma_0}}

\usepackage{ulem}

\begin{document}
\title{Splitting  the local Hilbert space:
MPS-based approach to large local dimensions}
\author{Naushad Ahmad Kamar}
\author{Mohammad Maghrebi}
\affiliation{Department of Physics and Astronomy, Michigan State University, East Lansing, MI 48823, USA}
\begin{abstract}
A large, or even infinite, local Hilbert space dimension poses a significant computational challenge for simulating quantum systems. In this work, we present a matrix product state (MPS)-based method for simulating one-dimensional quantum systems with a large local Hilbert space dimension, an example being bosonic systems with a large on-site population. To this end, we \textit{split} the local Hilbert space corresponding to one site into two sites, each with a smaller Hilbert space dimension. 
An advantage of this method is that it can be easily  integrated into MPS-based techniques such as time-dependent variational principle (TDVP) without changing their standard algorithmic structure.
Here, we implement our method using the TDVP to simulate the dynamics of the spin-boson model, a prototypical model of a spin interacting with a large bath of bosonic modes. We benchmark our method against and find excellent agreement with previous studies. 
\end{abstract}
\maketitle
\section{Introduction}
Characterizing the interaction between the bosonic modes and electronic or spin degrees of freedom is essential for understanding properties of materials~\cite{Haule_RMP_correlated_Materials,yonemitsu2008theory}, including superconductivity~\cite{marsiglio_electron_phonon_superconductivity}. A well-known example is the effect of electron-phonon coupling on the mass of electrons, which leads to the emergence of quasi-particles known as polarons~\cite{Feynman_Polaron}.
On the experimental front, circuit QED~\cite{Diaz_Ultra_Strong_Spin_Boson_model,magazzu2018probing,yoshihara2017superconducting,mirhosseini2019cavity} and trapped ions~\cite{Porras2008,Lemmer_2018}  
among others, provide highly controlled platforms for simulating a broad range of models of interest which also involve bosonic degrees of freedom with tunable coupling.  
A fundamental goal is to design perfect qubits in these platforms; however, in practice, such qubits are unavoidably coupled with the surrounding environment, which is often considered to be bosonic. 

The infinite local Hilbert space dimension of the bath, due to its bosonic nature, presents a significant numerical challenge; an exact diagonalization, even for small systems, would be difficult unless the bosonic population is low, 
in contrast with spin-1/2 or fermionic chains.
To cure this problem, Zhang et al.~\cite{Steven_White_Optimal_Basis} used the largest relevant eigenvalues and corresponding eigenvectors of the local density matrix to identify an effective local Hilbert space dimension that is smaller than the original one. In general, the local density matrix has $d_b$ eigenvalues with $d_b$ the original local Hilbert space dimension. However, in the ground state, these eigenvalues decrease rapidly; this allows for an approximation of the local density matrix through an optimal local Hilbert space with dimension $d_o \ll d_b$.
This method is called local-basis optimization, and the corresponding space is the optimal bosonic basis.
Various techniques~\cite{Brockt_Optimal_Bosonic_Basis, Guo_Spin_Boson_Model,Alex_Chin_Spin_Boson2016} that combine local basis optimization and MPS~\cite{schollwock_MPS} based methods such as time-evolving block decimation (TEBD)~\cite{Vidal_TEBD}, variational matrix product states (VMPS)~\cite{schollwock_MPS}, and TDVP~\cite{Haegeman_TDVP_Original,Haegeman_TDVP_Simple} have been utilized to investigate the ground state and dynamics of quantum systems that involve bosonic degrees of freedom. 
However, the local basis optimization changes the standard form of  VMPS~\cite{Guo_Spin_Boson_Model}, TEBD~\cite{Brockt_Optimal_Bosonic_Basis}, and TDVP~\cite{Alex_Chin_Spin_Boson2016}, and modify their algorithmic structure. For example, in VMPS, TEBD, and TDVP methods, one optimizes the MPS and the matrix corresponding to the orthogonality center of the MPS. However, introducing an optimal bosonic basis,  one should also optimize the local Hilbert space~\cite{Guo_Spin_Boson_Model,Brockt_Optimal_Bosonic_Basis,Alex_Chin_Spin_Boson2016} which drastically changes the structure of these MPS-based methods.

In this paper, we propose a simple method to treat a large local Hilbert space dimension without truncating the local density matrix, which preserves the algorithmic structure of VMPS and TDVP techniques. We exploit the sparsity of the Hamiltonian's local matrix product operator (MPO)~\cite{schollwock_MPS} and split the original local Hilbert space into two smaller ones using a matrix decomposition method, specifically the singular value decomposition. Upon splitting, the system doubles in linear size, but the local Hilbert space dimension reduces to $\sqrt d_b$. 
We apply our proposed method to the spin-boson model ~\cite{Leggett_spin_boson_model}, which describes the dynamics of a spin-1/2 strongly coupled to an infinite number of bosonic degrees of freedom---this prototypical model emerges in a variety of quantum systems~\cite{Diaz_Ultra_Strong_Spin_Boson_model,magazzu2018probing,yoshihara2017superconducting,mirhosseini2019cavity,Porras2008,Lemmer_2018}. Specifically, we simulate the dynamics by incorporating our method into the TDVP.  

The structure of the paper is as follows: In \cref{sec:model}, we introduce the spin-boson model and present a mapping to a short-range semi-infinite chain suitable for numerical simulation. In \cref{sec:Methods}, we briefly explain the standard MPS approach, and then introduce our main method. We provide numerical results benchmarking our method in \cref{sec:results}, and finally conclude and discuss future directions in \cref{sec:conclusion}. We provide further details of the MPO decomposition in \cref{sec:appendix}.
\section{Model}\label{sec:model}
We consider a  two-level system $S$, coupled with an infinite number of non-interacting bosons, famously known as the spin-boson model~\cite{Leggett_spin_boson_model}. We describe the system-bath coupling via the Hamiltonian
\begin{equation} \label{eq:eq0a}
\begin{split}
H&=H_S+H_B+H_{SB},
\end{split}
\end{equation}
where the Hamiltonians $H_S$, $H_B$, and $H_{SB}$ describe the system, bath, and the linear coupling between the system and the bath, respectively, 
\begin{equation}\label{eq:SBM1}
\begin{split}
H_S&=-\frac{\Delta}{2} \sigma^x,\\
H_B&=\sum_k\omega_ka_k^\dagger a_k,\\
H_{SB}&=\frac{\sigma^z}{2}\sum_k \lambda_k (a_k^\dagger+a_k).
\end{split}
\end{equation}
\begin{figure}
\begin{center}
\includegraphics [ scale=0.45]
{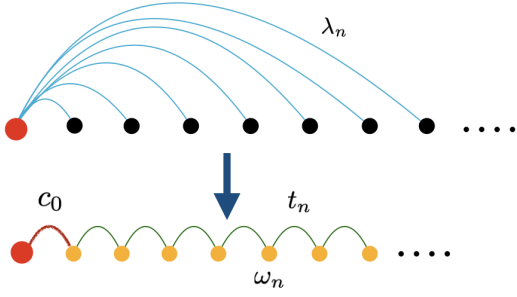}
\end{center} 
\caption{\label{fig:model} (color online) Two lattice representations of the spin-boson model. The top panel shows the spin-boson model introduced in \cref{eq:SBM1}. The large (red) circle represents the spin, and the small (black) circles indicate the bosonic modes. The spin-boson coupling $\lambda_n$ is denoted by the (blue) curves. In the lower panel, the spin-boson model is shown in the transformed basis given by  \cref{eq:chain_mapping}. The small (orange) circles represent the bosonic modes in the transformed basis that defines a tight binding model with site-dependent energy $\omega_n$ and tunneling amplitude $t_n$. The spin interacts directly with the first bosonic site with a strength $c_0$.}
\end{figure}
The effective coupling between the spin and the bath depends on $\omega_k$ and $\lambda_k$, and is fully characterized by the spectral function $J(\omega)$ defined as 
\begin{equation}\label{eq:spectral}
\begin{split}
J(\omega)&=\pi\sum_k \lambda_k^2\delta(\omega-\omega_k).
\end{split}
\end{equation}
Depending on its form, the spectral function could describe a wide range of different qualitative behavior. A representative class of quantum baths are described by the spectral function 
\begin{equation}\label{eq:spectral1}
\begin{split}
J(\omega)&=2\pi\alpha \omega^s\omega_c^{1-s}\Theta(\omega_c-\omega),
\end{split}
\end{equation}
corresponding to an Ohmic bath with $s=1$ and sub-(super-)Ohmic baths where $s<1$ ($s>1$). 
The parameters $\alpha$ and $\omega_c$ characterize the coupling strength and the frequency cutoff of the bath, respectively.
For an  Ohmic bath,  $s=1$, this model exhibits a quantum phase transition from a delocalized to localized state at $\alpha\simeq 1+\mathcal O(\Delta/\omega_c)$~\cite{Leggett_spin_boson_model,Guo_Spin_Boson_Model,Alex_Chin_Spin_Boson2016}. Similar quantum phase transitions occur for the sub-Ohmic bath \cite{Ralf_Bulla_NRG_Spin_Boson_Model,Bulla_Spin_Boson_Model_Sub_Ohmic_RG}.

The spin-boson model in \cref{eq:eq0a} couples the spin to all the bosonic modes, mimicking a kind of long-range interaction, as depicted in the upper panel of Fig.~\ref{fig:model}; this makes a simulation based on matrix product states rather expensive. However, by using an appropriate basis transformation of the bosonic local operators $a(a^\dagger)$, this model can be mapped to a nearest-neighbor Hamiltonian as~\cite{Ralf_Bulla_NRG_Spin_Boson_Model,Chain_Mapping_of_Spin_Boson_Model+Alex_Chin}  
\begin{equation}\label{eq:chain_mapping}
\begin{split}
H=&-\frac{\Delta}{2} \sigma^x+c_0\sigma^z(b_0+b_0^\dagger)+\sum_{n=0}^{L}\omega_n b_n^\dagger b_n\\
&+\sum_{n=0}^{L-1} t_n(b_n^\dagger b_{n+1}+h.c),
\end{split}
\end{equation}
where $b_n$s define the bosonic operators in the new basis, and $\omega_n$, $t_n$, and $c_0$ denote the local energy, site-dependent tunneling amplitude, and the coupling between the spin and the first site in the bath in the new basis; see also the lower panel of Fig.~\ref{fig:model}. The above coefficients can be computed exactly and are given by \cite{Chain_Mapping_of_Spin_Boson_Model+Alex_Chin} 
\begin{equation}\label{eq:chain_parameters}
\begin{split}
\omega_n&=\frac{\omega_c}{2}\Big(1+\frac{s^2}{(s+2n)(2+s+2n)}\Big),\\
t_n&=\frac{\omega_c(1+n)(1+s+n)}{(s+2+2n)(3+s+3n)}\sqrt{\frac{3+s+2n}{1+s+2n}},\\
c_0&=\sqrt{\frac{\alpha}{2(1+s)}}\omega_c.
\end{split}
\end{equation}
While being local, this model comprises bosonic modes whose population can be large, thus posing a challenge for numerical simulation. In the next section,  we introduce an MPO decomposition to split a large local Hilbert space into smaller ones. Combined with MPS-based methods, this allows us to simulate systems with a large on-site bosonic population.
\begin{figure}[t]
\begin{center}
\includegraphics [ scale=0.23]
{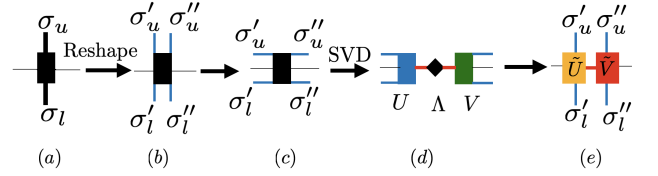}
\end{center} 
\caption{\label{fig:Operator} (color online) Schematics of the MPO decomposition. 
In panel (a), an MPO is shown in the original basis $|\sigma_u\rangle\langle \sigma_l|$. In panels (b, c), we reshape the MPO matrix as $W^{\sigma_u}_{\sigma_l}(w_{i-1},w_i) \to W^{\sigma'_u \sigma''_u}_{\sigma'_l \sigma''_l}(w_{i-1},w_i) \to W[{\sigma'_u \sigma'_l w_{i-1}},{\sigma''_l\sigma''_u}w_i]$ first in the split basis  spanned by $|\sigma'\rangle$ and $|\sigma''\rangle$, and then into a matrix form, where $w_i$ is index of the MPO bond dimension.  
In panel (d), we split this matrix using an SVD resulting in the left singular matrix $U$, singular values $\Lambda$, and the right singular matrix $V$. Finally, in panel (e), we absorb the singular values in the left and right singular matrices as $\tilde{U}=U\sqrt\Lambda$ and $\tilde{V}=\sqrt\Lambda V$. In this process, we have split a given site's MPO $W$ into two sites with the MPOs $\tilde U$ and $\tilde V$. }
\end{figure}
\section{Method}\label{sec:Methods}
In this section, we briefly introduce the MPS and MPO~\cite{schollwock_MPS} in order to simulate the spin-boson model. The state of the spin-boson model in the MPS language is given by 
\begin{equation}\label{eq:eq1}
|\psi\rangle=\sum_{\sigma_0,\sigma_1,..,\sigma_L}A^{\sigma_0}[0]A^{\sigma_1}[1]...A^{\sigma_L}[L]|\sigma_0,\sigma_1,..,\sigma_L\rangle,
\end{equation}
where $\sigma_0$ runs from $1$ to $d$ and $\sigma_{1,2,..,L}$ run from $1$ to $d_b$, where $d$ and $d_b$ are the local Hilbert space dimension of the spin and bosons, respectively. The size of the $A$ matrices bounds the maximum entanglement that can exist in the system.
In a similar fashion, an operator can also be defined using a product of operators known as MPO. In general, the MPO of a given Hamiltonian can be constructed as
\begin{equation} \label{eq:MPO1}
\begin{split}
H=\sum_{\sigma_{u/l0},\cdots, \sigma_{u/lL}}  &W^{\sigma_{u0}}_{\sigma_{l0}}[0]W^{\sigma_{u1}}_{\sigma_{l1}}[1] \cdots W^{\sigma_{uL}}_{\sigma_{lL}}[L] \\ 
\times \,\, &|\sigma_{u0},\sigma_{u1}...,\sigma_{uL}\rangle\langle \sigma_{l0},\sigma_{l1}...,\sigma_{lL}|\,,
\end{split}
\end{equation}
where $\sigma_{u/l n}$ denotes the ket/bra indices on site $n$. 
For the spin-boson model, the  $W$ matrices in the MPO are explicitly given by
\begin{equation}\label{eq:MPO}
\begin{split}
W[0]&=
\begin{pmatrix}
I_s&\sigma^z&0&0&-\frac{\Delta}{2}\sigma^x\\
\end{pmatrix},\\
W[1]&=
\begin{pmatrix}
I_b&0&b^\dagger&b&\omega_0n_b\\
0&0&0&0&c_0(b^\dagger+b)\\
0&0&0&0&t_0 b\\
0&0&0&0&t_0 b^\dagger\\
0&0&0&0&I_b
\end{pmatrix},\\
W[1<n<L]&=
\begin{pmatrix}
I_b&0&b^\dagger&b&\omega_{n-1}n_b\\
0&0&0&0&0\\
0&0&0&0&t_{n-1} b\\
0&0&0&0&t_{n-1} b^\dagger\\
0&0&0&0&I_b
\end{pmatrix},\\
W[L]&=
\begin{pmatrix}
\omega_{L-1}n_b\\
0\\
t_{L-1} b\\
t_{L-1} b^\dagger\\
I_b
\end{pmatrix}.\\
\end{split}
\end{equation}
Here, $I_s$ and $I_b$ refer to the identity operators for the spin and the bath, respectively; $I_s$ is a $2\times2$ matrix for the spin-1/2 while $I_b$ is a $d_b\times d_b$ matrix. We have also defined the local number operator on a given site in the bath as $n_b = b^\dagger b$.  The MPO of the spin-boson model, as described in \cref{eq:MPO}, is a $5\times5$ matrix of operators defined in the local Hilbert space.
\begin{figure}
\begin{center}
\includegraphics [ scale=0.4]
{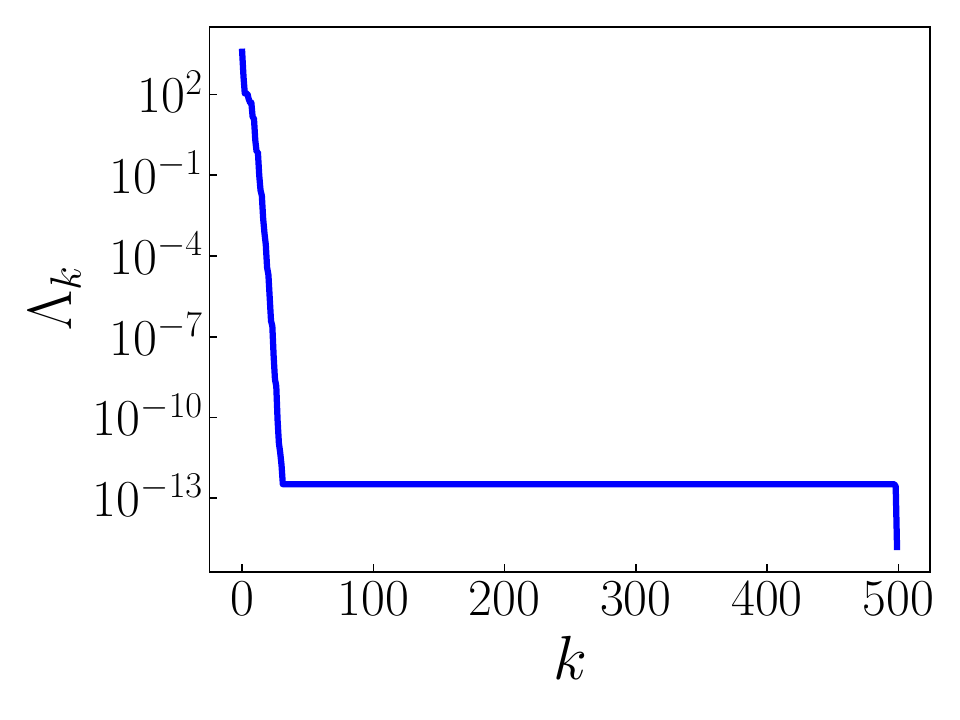}
\end{center} 
\caption{\label{fig:MPO_SCH} 
(color online) Singular values of an MPO of the spin-boson model at site $n=2$  on a semi-log scale for $d_b=100$, $\alpha=1.0$, $\omega_c=1.0$, $\Delta=0.1$, and $\omega_c=1$. The MPO is shown on site $n=2$; we find similar behavior on all sites. The singular value decrease exponentially with $k$, and are effectively zero beyond $k=29$.The MPO bond dimension in the split basis is an order of magnitude smaller than its maximum value of $X_Wd_b =500$. }
\end{figure}
In the MPS-based methods such as VMPS and TDVP, the computational complexity scales with the $3^{\rm rd}$ power of the local Hilbert space dimension $d_b$. Therefore, for a large $d_b$, these methods become computationally expensive. To circumvent this problem, we break up, or split, the local Hilbert space $\cal H$ into two Hilbert spaces, $\cal H=\cal H'\otimes \cal H''$, each with a smaller dimension.
A basis state $|\sigma\rangle$ in $\cal H$ can be then expressed as a product state 
\begin{align} \label{eq:eq2}
|\sigma\rangle&=|\sigma{'} \rangle \otimes |\sigma{''}\rangle,
\end{align}
where $|\sigma{'} \rangle$ and $|\sigma{''} \rangle$ are defined in  $\cal H'$ and $\cal H''$, respectively, and the corresponding indices $\sigma',\sigma''$ run 
from $1$ to $\sqrt {d_b}$. There are of course many ways to split the original basis; here, we choose a particular factorization scheme where 
\begin{align}
\sigma=\sqrt{d_b}(\sigma'-1)+\sigma''.
\label{eq:spllitting}
\end{align}
Such splitting scheme can easily be  implemented using, for example, the Numpy's reshape library.   
Next, the state $|\psi\rangle$ in \cref{eq:eq1} can be recast in the new basis as 
\begin{align} \label{eq:eq10}
|\psi\rangle=\sum_{\sigma_0,\sigma{'_1}, \sigma{''_1},\cdots,\sigma{'_L}, \sigma{''_L}} A^{\sigma_0}[0]\tilde{A}^{\sigma{'_1}}[1]\tilde{A}^{\sigma{''_1}}[2]\cdots\\\nonumber \times \tilde{A}^{\sigma{'_L}}[2L-1]\tilde{A}^{\sigma{''_L}}[2L]|\sigma_0,\sigma{'_1},\sigma{''_1},\cdots,\sigma{'_L},\sigma{''_L}\rangle,
\end{align}
where we have introduced the new matrices $\tilde A$ now spanning sites 1 to $2L$. Each site being split into two, the linear size of the chain is doubled.

The local MPO  matrices $W$ can also be expressed in the new basis by using singular value decomposition (SVD) as 
\begin{align} \label{eq:MPO2}
W&=U\Lambda V,
\end{align}
where the matrix $U$ and $V$ are defined in the new basis spanned by $|\sigma'\rangle$ and $|\sigma''\rangle$, respectively. We leave the technical details to \cref{sec:appendix}; for a schematic explanation, see \cref{fig:Operator}. For simplicity, we can absorb the diagonal matrix $\Lambda$ containing the singular values of the SVD into the definition of the $U$ and $V$ matrices as
\begin{align}
\tilde{U}=U\sqrt\Lambda, \quad 
\tilde{V}=\sqrt\Lambda V,
\end{align}
upon which \cref{eq:MPO2} simply becomes 
\begin{align} \label{eq:MPO3}
W&=\tilde U \tilde V.
\end{align}
The column (row) dimension of $\tilde U$ $(\tilde V)$ is $d_b X_W$ where $X_W$ is the MPO bond dimension before splitting; e.g., $X_W=5$ for the spin-boson model. 
In practice, however, the effective MPO bond dimension in the split basis could be taken to be much smaller as the singular values $\Lambda_k$ of the matrix $\Lambda$ decay rather quickly with the index $k$. 
As a representative example, we consider the spin-boson model with $d_b=100$, $\omega_c=1$, $\Delta=0.1$ and $\alpha=1.0$, and show 
$\Lambda_k$ in descending order in \cref{fig:MPO_SCH}. We observe that $\Lambda_k$ rapidly decreases and is practically vanishing beyond $k=29$, therefore effectively $29$ instead of $X_W d_b =500$; the row (column) dimension of $\tilde U$ $(\tilde V)$ is still $X_W =5$ since the MPO structure has not changed on the original bonds before splitting.
MPS and MPO play a crucial role in MPS-based techniques such as VMPS, TDVP, TEBD, and MPO-MPS time evolution~\cite{METTS_Stoudenmire,Zalatel_Long_range_Time_evolution}. In our approach, we have split the original local Hilbert space into local Hilbert spaces with smaller dimensions while leaving the algorithmic structure of the MPS intact in the new basis.
The only difference is that the MPS is now optimized with the smaller local Hilbert space dimension of $\sqrt d_b$ in the split basis. While the computational complexity scales as $\mathcal O(d_b^3)$ in the original basis,  it scales as  $\mathcal O(2d_b^{3/2})$ in the new basis, where the factor of $2$ is due to the system size being doubled. 
We thus expect that the MPS-based methods in the split basis feature a speedup by a factor of the order of $\mathcal O(d_b^{3/2}/2)$ compared to the old basis; this is a massive speedup for large local Hilbert space dimensions. 
In the next section, we use the spin-boson model as a testbed for our method.
\begin{figure}
\begin{center}
\includegraphics [ scale=0.4]
{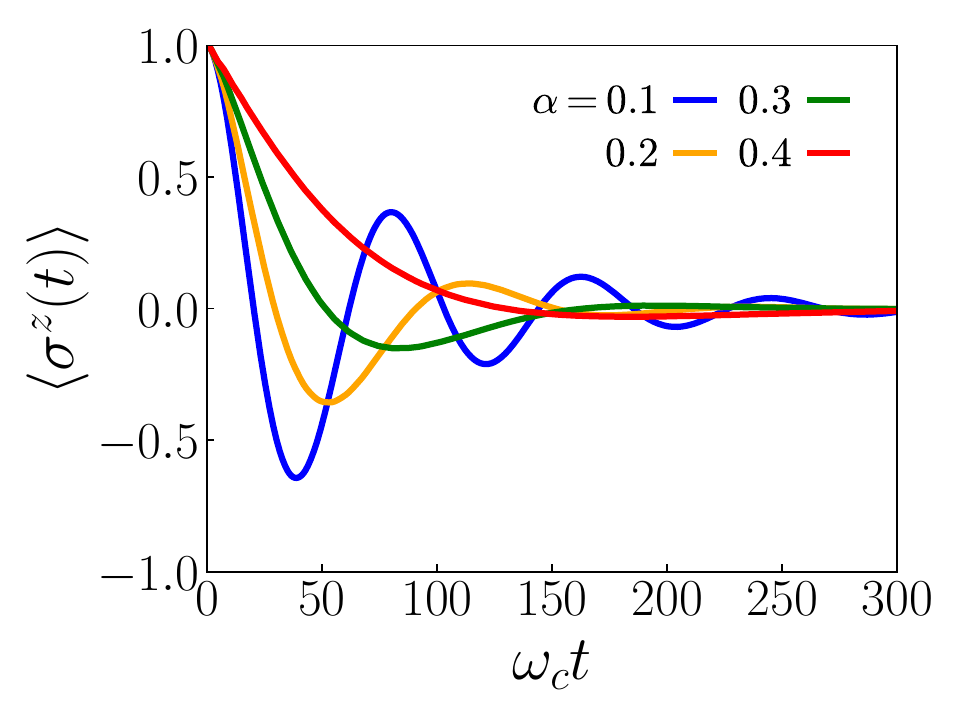}
\hspace*{0.5cm}\includegraphics [ scale=0.37]
{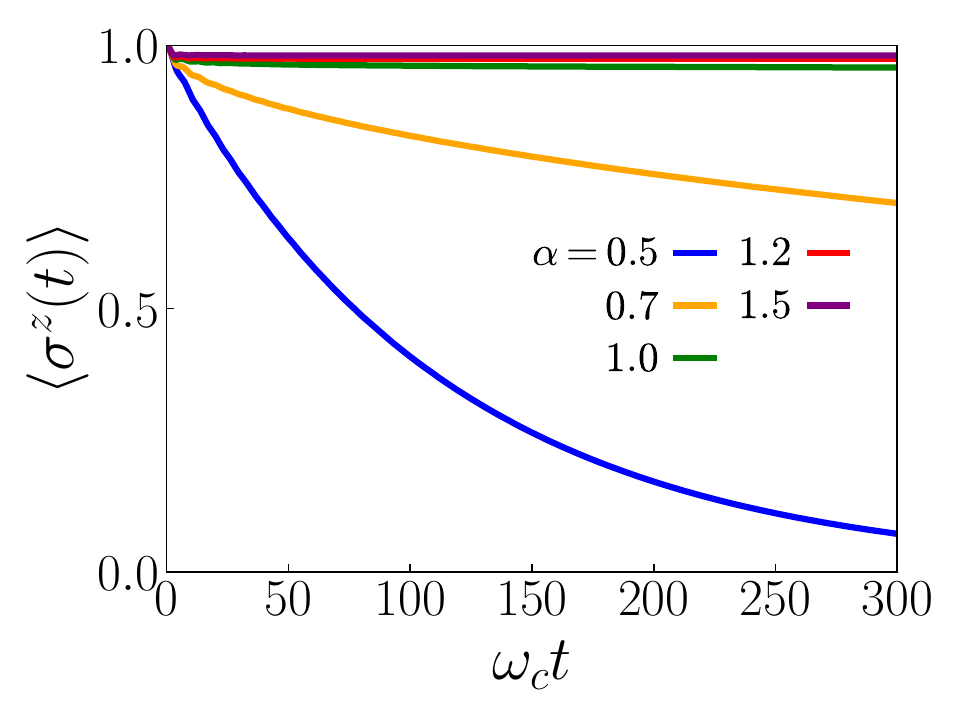}
\end{center} 
\caption{\label{fig:magnetization_Ohmic} (color online) Magnetization $\langle\sigma^z(t)\rangle$ as a function of time in the presence of an Ohmic bath, $s=1$, and at different values of $\alpha$; we have taken  $\Delta=0.1$ and $\omega_c =1.0$. The upper panel depicts  $\langle\sigma^z(t)\rangle$  for $\alpha=0.1-0.4$. The spin shows coherent damping as a function of time, with the frequency of oscillations decreasing with $\alpha$. The lower panel depicts  $\langle\sigma^z(t)\rangle$ for $\alpha=0.5, 0.7, 1.0, 1.2$, and $1.5$. The spin shows incoherent damping as a function of time for $0.5 \le \alpha<1$. At and beyond the critical point $\alpha_c=1.0$, 
the dynamics is frozen close to $|\uparrow\rangle$, signalling a quantum phase transition from a delocalized to a localized  phase.  }
\end{figure}
\begin{figure}
\begin{center}
\includegraphics [ scale=0.4]
{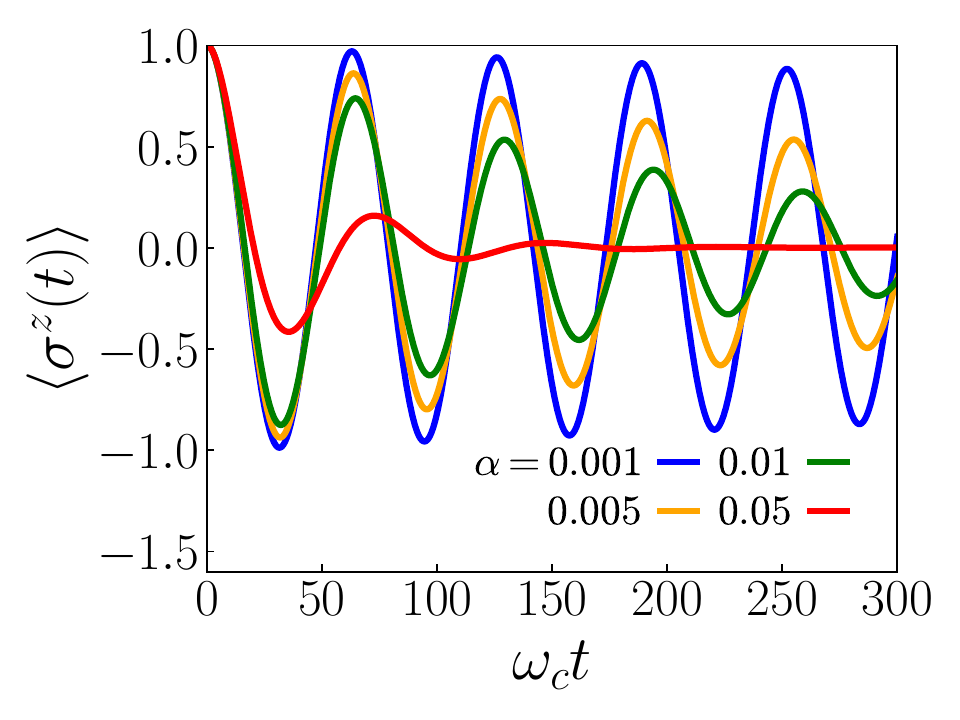}
\includegraphics [ scale=0.4]
{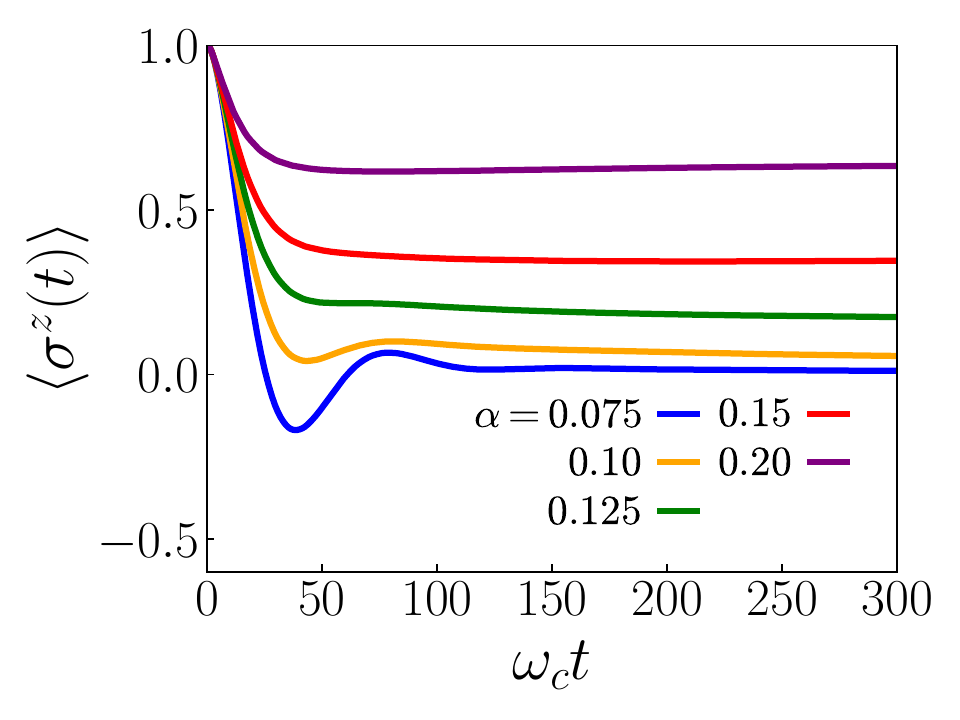}
\end{center} 
\caption{\label{fig:magnetization_SubOhmic} (color online) Magnetization $\langle\sigma^z(t)\rangle$ as a function of time in the presence of a sub-Ohmic bath with $s=0.5$ and  at different values of $\alpha$; we have taken $\Delta=0.1$ and $\omega_c =1$. The upper panel depicts the $\langle\sigma^z(t)\rangle$  for $\alpha=0.001-0.05$, which displays coherent damping as a function of time; the oscillation frequency decreases with $\alpha$ similar to the Ohmic case. The lower panel depicts $\langle\sigma^z(t)\rangle$ for $\alpha=0.075-0.20$. 
The spin exhibits overdamped dynamics for $\alpha>0.1$ before it enters the localized phase around $\alpha=0.125$. }
\end{figure}
\section{Results}\label{sec:results}
In this section, we apply our method to the spin-boson model and specifically study the dynamics of the spin. We start from an initial state at $t=0$ where the spin is in the $|\uparrow\rangle$ state (in the $\sigma^z$ basis), and the bosonic modes are in their vacuum state, 
\begin{align}
|\psi(0)\rangle=|\uparrow\rangle\otimes|0\rangle\otimes \cdots\otimes|0\rangle,
\end{align}
where $b|0\rangle=0$ and $\sigma^z|\!\!\uparrow\rangle=|\uparrow\rangle$.
We are mainly interested in the time evolution of magnetization defined by $\langle\sigma^z(t)\rangle=\langle\psi(t)|\sigma^z|\psi(t)\rangle$ 
where 
\begin{align}\label{eq:sch}
|\psi(t)\rangle=e^{-i H t}|\psi(0)\rangle.
\end{align}
In order to compute $|\psi(t)\rangle$, we employ the TDVP algorithm in the new basis. We fix the interaction parameters at $\Delta=0.1$, $d_b=100$, $\omega_c=1$, $L=100$, and take the MPS bond dimension $\chi= 5$. We study the dynamics for both  Ohmic ($s=1$) and sub-Ohmic (with $s=0.5$) baths. 
\begin{figure}[t]
\begin{center}
\includegraphics [ scale=0.4]
{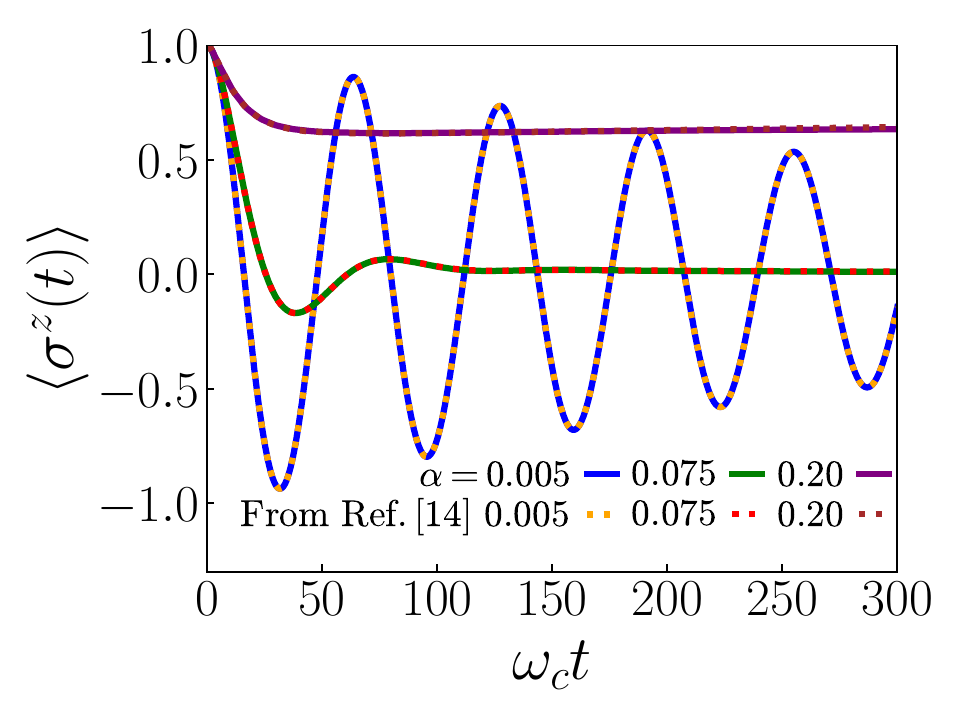}
\end{center} 
\caption{\label{fig:magnetization_SubOhmic_Alex} (color online) Magnetization $\langle\sigma^z(t)\rangle$ as a function of time in the presence of the sub-Ohmic bath at different values of $\alpha$, and with $\Delta=0.1$, $\omega_c=1.0$ and $s=0.5$. Our results (the solid lines) are contrasted against the data 
taken from Ref.~\cite{Alex_Chin_Spin_Boson2016} (dotted lines), which are obtained using TDVP combined with the optimal bosonic basis. The excellent agreement with this data is a nontrivial check of our method.}
\end{figure}
In \cref{fig:magnetization_Ohmic}, we depict $\langle\sigma^z(t)\rangle$ as a function of time for different values of the interaction parameters in the range $\alpha=0.1-1.5$ and for $s=1$. For $\alpha=0.1-0.4$, we find that the dynamics 
is underdamped; see the upper panel of \cref{fig:magnetization_Ohmic}. The frequency of oscillations decreases while the damping rate increases with $\alpha$, in harmony with the previous studies~\cite{Leggett_spin_boson_model,Peter_Stochastic_Schrodinger_Equation,Shapourian_2016,kamar2023spin}. Specifically, the oscillation frequency is renormalized by the spin-bath coupling $\alpha$ as $\Delta_r=\Delta(\Delta/\omega_c)^{\frac{\alpha}{1-\alpha}}$~\cite{Leggett_spin_boson_model,Peter_Stochastic_Schrodinger_Equation,Shapourian_2016,kamar2023spin};  we have verified that our results are in quantitative agreement with this equation. For $\alpha=0.5, 0.7$, we observe that  $\langle\sigma^z(t)\rangle$
decays exponentially to zero, a behavior which persists in the range $0.5\le\alpha<1$~\cite{Leggett_spin_boson_model,Alex_Chin_Spin_Boson2016}. At or above the critical point $\alpha_c=1.0$, the magnetization $\langle\sigma^z(t)\rangle$ barely decays and is localized in the $|\uparrow\rangle$ state; see the lower panel of \cref{fig:magnetization_Ohmic}.
This signals a quantum phase transition from a delocalized to a localized state, again consistent with the previous results~\cite{Leggett_spin_boson_model,Ralf_Bulla_NRG_Spin_Boson_Model,Alex_Chin_Spin_Boson2016}. 

As another example, we consider the dynamics of the spin coupled to a sub-Ohmic bath. In \cref{fig:magnetization_SubOhmic}, we show $\langle\sigma^z(t)\rangle$ as a function of time in the presence of a sub-Ohmic bath with $s=0.5$ and for $\alpha=0.005$-$0.20$. Again, we can identify the underdamped regime (upper panel) as well as the overdamped and localized regimes (lower panel).
We find that $\langle \sigma^z\rangle$  saturates to a nonzero value beyond $\alpha=0.125$, signaling a quantum phase transition to a localized phase, consistent with Ref.~\cite{Alex_Chin_Spin_Boson2016}. 
Finally, in \cref{fig:magnetization_SubOhmic_Alex}, we compare our numerical results with those presented in Ref.~\cite{Alex_Chin_Spin_Boson2016} for $s=0.5$ and different values of $\alpha$. We find that our numerical results exactly match the data presented in Ref.~\cite{Alex_Chin_Spin_Boson2016}, thus providing a nontrivial check of the accuracy and efficiency of our method. An advantage of our method is its simple structure which can be easily integrated into the standard MPS-based methods.
\section{Conclusion and perspective}\label{sec:conclusion}
We have proposed a simple computational approach to simulate systems involving a large local Hilbert space dimension. Our method is based on splitting a large local Hilbert space into two sites with a smaller dimension. We have shown that our approach correctly reproduces the results obtained from the TDVP combined with the local basis optimization for the spin-boson model~\cite{Alex_Chin_Spin_Boson2016}. Our method has the advantage that it does not change the algorithmic structure of MPS-based methods, in contrast with MPS approaches that utilize the local basis optimization~\cite{Brockt_Optimal_Bosonic_Basis,Guo_Spin_Boson_Model,Alex_Chin_Spin_Boson2016}. 

Our numerical method becomes even more vital in simulating bosonic systems described by a mixed state either at finite temperature or in open quantum systems, e.g., in systems described by the Lindblad master equation. In these scenarios, one generally vectorizes the density matrix in order to bring it to a form that can be represented in the MPS form; however, the local Hilbert space dimension becomes the square of the original local Hilbert space dimension, which could pose a challenge for numerical simulations (see also \cite{wolff2020numerical}).
Our approach provides a formidable alternative to simulate systems described by a large local Hilbert space dimension. 

In this work, we have proposed a method to treat large local Hilbert spaces by splitting them into smaller ones. It is worthwhile extending this idea to a large \textit{bond dimension} where a local MPS is decomposed into two or more matrices with a smaller bond dimension, leading to ladder-like lattices. Such an approach could result in more efficient MPS-based calculations where the original bond dimension is large. 
\begin{acknowledgments}
We thank Alex Chin for useful discussions. This work is supported by the Air Force Office of Scientific Research (AFOSR) under the award number FA9550-20-1-0073. We also acknowledge support from the National Science Foundation under the NSF CAREER Award (DMR2142866), as well as the NSF grants DMR1912799 and PHY2112893. 
\end{acknowledgments}
\appendix 
\section{MPO Splitting in $\sigma'$ and $\sigma''$ basis}\label{sec:appendix}
In this Appendix, we provide further details for the decomposition of a local MPO in terms of the two MPOs with smaller local Hilbert space dimensions.
We first split $|\sigma\rangle$ into $|\sigma{'} \rangle$ and $|\sigma{''} \rangle$ as 
\begin{align} \label{eq:eq2}
|\sigma\rangle&=|\sigma{'} \rangle \otimes |\sigma{''}\rangle,
\end{align}
where $\sigma{'}$ and $\sigma{''}$ runs from $1$ to $\sqrt d_b$.
We can express the local MPO matrix $W$ as a four-dimensional array of size $d_b\times d_b\times X_W \times X_W$, where $X_W$ is the MPO bond dimension. We express the corresponding array elements as 
\begin{align}\label{eq:eq4}
W[\sigma_u,\sigma_l,w_{i-1},w_i]=W^{\sigma_u}_{\sigma_l}(w_{i-1},w_i),
\end{align}
where $w_i$ runs from $1$ to $X_W$, and in an abuse of notation we used the same symbol $W$ to denote the array. 
Splitting the local basis as in \cref{eq:eq2}, the above array can be recast as a six-dimensional array in the new basis:
\begin{align}\label{eq:eq5}
W[\sigma'_u,\sigma''_u,\sigma'_l,\sigma''_l,w_{i-1},w_i]=W^{\sigma'_u,\sigma''_u}_{\sigma'_l,\sigma''_l}(w_{i-1},w_i).
\end{align}
We can reshape $W$ again to bring it into the form 
\begin{equation}
W[\sigma'_u,\sigma''_u,\sigma'_l,\sigma''_l,w_{i-1},w_i] \to W[w_{i-1},\sigma'_u,\sigma'_l,\sigma''_u,\sigma''_l,w_i]\,.
\end{equation}
Finally we can express $W$ in matrix form as
\begin{equation}\label{eq:eq6}
W[w_{i-1},\sigma'_u,\sigma'_l,\sigma''_u,\sigma''_l,w_i] \to W[w_{i-1}\sigma'_u\sigma'_l,\sigma''_u\sigma''_lw_i]\,.
\end{equation}
The MPO $W$ can be then factorized using the SVD as 
\begin{align}\label{eq:eq7}
W[w_{i-1}\sigma'_u\sigma'_l,\sigma''_u\sigma''_lw_i]=\sum_{k=1}^{X_Wd_b} U[w_{i-1}\sigma'_u\sigma'_l,k]\Lambda_k V[k,\sigma''_u\sigma''_lw_i].
\end{align}
In the above equation $k$ runs from 1 to $X_Wd_b$; however, in practice, $\Lambda_k$ decays rapidly with $k$ and most of the singular values are zeros, and we can set the upper limit to some $k_{\text{eff}} < X_W d_b$. 

Finally, the above equation can be written as 
\begin{align}\label{eq:eq8}
\begin{split}
W[w_{i-1}\sigma'_u\sigma'_l,\sigma''_u\sigma''_lw_i]&=\sum_{k=1}^{k_{\mathrm{eff}
}} U[w_{i-1}\sigma'_u\sigma'_l,k]\Lambda_k V[k,\sigma''_u\sigma''_lw_i],\\
&=\sum_{k=1}^{k_{\mathrm{eff}}}\widetilde{U}[w_{i-1}\sigma'_u\sigma'_l,k]\widetilde{V}[k,\sigma''_u\sigma''_lw_i],\\
\end{split}
\end{align}
where we have defined
\begin{align}
\begin{split}
\widetilde{U}[w_{i-1}\sigma'_u\sigma'_l,k]&=U[w_{i-1}\sigma'_u\sigma'_l,k]\sqrt{\Lambda_k}\,,\\
\widetilde{V}[k,\sigma''_u\sigma''_lw_i]&=\sqrt{\Lambda_k} V[k,\sigma''_u\sigma''_lw_i]\,.
\end{split}
\end{align}
The matrices $\widetilde{U}[w_{i-1}\sigma'_u\sigma'_l,k]$ and $\widetilde{V}[k,\sigma''_u\sigma''_lw_i]$ in the above equation can again be reshaped as 
\begin{align}\label{eq:eq8}
\widetilde{U}[w_{i-1}\sigma'_u\sigma'_l,k]&\to \widetilde{U}[w_{i-1},\sigma'_u,\sigma'_l,k]\to \widetilde{U}[\sigma'_u,\sigma'_l,w_{i-1},k]\nonumber,\\
\widetilde{V}[k,\sigma''_u\sigma''_lw_i]&\to \widetilde{V}[k,\sigma''_u,\sigma''_l,w_i]\to \widetilde{V}[\sigma''_u,\sigma''_l,k,w_i],
\end{align}
where $\widetilde{U}$ and $\widetilde{V}$ represent the MPO in $|\sigma'\rangle$ and $|\sigma''\rangle$ basis, respectively. In the new basis the MPO of the split sites can be then expressed as 
\begin{align} 
W&=\widetilde{U}\widetilde{V}\,.
\end{align}
A schematic figure summarizing the above steps is illustrated in \cref{fig:Operator}.
\bibliography{Spin_Boson_ref}
\appendix
\end{document}